# pyspect: An Extensible Toolbox for Automatic Construction of Temporal Logic Trees via Reachability Analysis

Kaj Munhoz Arfvidsson[1], Loizos Hadjiloizou[2], Frank J. Jiang[1], Karl H. Johansson[1], Jonas Mårtensson[1]

*Abstract*—In this paper, we present pyspect, a Python toolbox that simplifies the use of reachability analysis for temporal logic problems. Currently, satisfying complex requirements in cyber-physical systems requires significant manual effort and domain expertise to develop the underlying reachability programs. This high development effort limits the broader adoption of reachability analysis for complex verification problems. To address this, pyspect provides a method-agnostic approach to performing reachability analysis for verifying a temporal logic specification via temporal logic trees (TLTs). It enables the specification of complex safety and liveness requirements using high-level logic formulations that are independent of any particular reachability technique or set representation. As a result, pyspect allows for the comparison of different reachability implementations, such as Hamilton-Jacobi and Hybrid Zonotope-based reachability analysis, for the same temporal logic specification. This design separates the concerns of implementation developers (who develop numerical procedures for reachability) and end-users (who write specifications). Through a simple vehicle example, we demonstrate how pyspect simplifies the synthesis of reachability programs, promotes specification reusability, and facilitates side-by-side comparisons of reachability techniques for complex tasks.

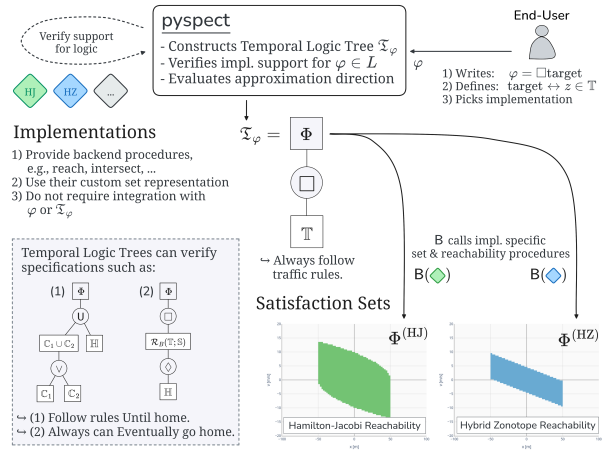

Fig. 1: Overview of the `pyspect` workflow. The end-user defines temporal logic specification $\varphi$. `pyspect` constructs a corresponding TLT which automatically generate a reachability program that verifies $\varphi$ using compatible implementations. With HJ and HZ backends, `pyspect` facilitates computation of $\Phi^{(HJ)}$ and $\Phi^{(HZ)}$, respectively.

## I. INTRODUCTION

Formal methods for cyber-physical systems (CPS) have gained significant attention in recent decades as these systems become ubiquitous in safety-critical applications. In particular, there is a rapidly growing trend to express requirements in temporal logic (e.g. linear temporal logic or signal temporal logic) due to their expressive power and formal rigor [1], [2]. Specifically, using temporal logic to encode invariance and liveness objectives can improve the interpretability of system design while also facilitating formal verification [3], [4].

However, applying temporal logic to CPS using traditional automata-based approaches poses several fundamental challenges. In particular, automata-based methods struggle in key three scenarios that arise in CPS applications: (1) when low-level, continuous dynamics need to be considered in the verification, (2) when the specification or environment changes during deployment, and (3) when not only a single control policy, but a satisfying set of control inputs need to be synthesized [5]. Challenges (1) and (2) can be treated through a variety of approaches that extend classical automata-based formal methods using abstractions to formalize a relation between a CPS's continuous state space to a finite transition system [6], [7] or sampling to improve the reactivity of the verification system [8], [9].

Recently, a new computational tool called temporal logic trees (TLT) was developed to address all three challenges. TLTs provide a structured approach to verify temporal logic specifications directly with reachability analysis, instead of using abstractions or sampling-based approaches [10], [11]. Using TLTs for formal verification, a temporal logic specification can be verified directly in the continuous state space of the CPS while accounting for its low-level, and possibly nonlinear, dynamics. Due to the modularity of TLTs, the verification result can be rapidly updated when the specification changes online. Moreover, since the TLT is constructed using state sets, the TLT can be used to synthesize the set of satisfying control policies [12]. However, in practice, building these trees requires deep expertise in both temporal logic and numerical methods for performing reachability analysis, posing a significant barrier to adoption and large-scale use in CPS applications.

In this paper, we address this gap through a carefully designed, extensible toolbox that integrates temporal logic tree construction with reachability analysis using a rigorous and accessible Python interface. We start by making the observation that many existing approaches to combining temporal logic with reachability analysis commonly require writing specialized problem definitions and solution methods that couple specification and algorithm, forcing users to adopt specific input languages, logic semantics, or reachability analysis approaches. This makes it difficult to interchange reachability frameworks or reuse a specification across different tools, even if they aim to solve the same verification problem. Each time a specification is written, it often calls for a re-formulation of the reachability problem, significantly slowing down the development of new applications. To alleviate this issue, we design the toolbox "Python Specification and Control with Temporal Logic Trees" (pyspect). pyspect is designed to be a method-agnostic tool for defining and verifying temporal logic specifications using any reachability analysis approach. This enables a clear separation of concerns between "what to verify" and "how to verify it." To illustrate this, we use pyspect to implement the verification of temporal logic specifications using both Hamilton-Jacobi reachability analysis [12] and hybrid zonotope-based reachability analysis [13]. By prioritizing modularity, clarity, and rigor in the interface design, pyspect lowers the barrier to developing new verification pipelines and accelerates the application of temporal logic to emerging safety-critical CPS domains. Specifically, this work contributes the following:

1) we decouple reachability problem definitions from their computational implementations and set representations;
2) we define an operational semantics for specification languages using TLT primitives, statically determining the approximation direction (over/under) of built TLTs and verifying their compatibility with solver backends;
3) and subsequently, we operationalize individual specifications, generating their correct reachability programs that execute on compatible solvers.

[1] K. Munhoz Arfvidsson, F. J. Jiang, K. H. Johansson and J. Mårtensson are with the Division of Decision and Control Systems, EECS, KTH Royal Institute of Technology, Malvinas väg 10, 10044 Stockholm, Sweden, email: {kajarf, frankji, kallej, jonas1}@kth.se. They are also affiliated with the Integrated Transport Research Lab and Digital Futures.

[2] L. Hadjiloizou is with the Division of Robotics, Perception and Learning, EECS, KTH Royal Institute of Technology, Lindstedtsvägen 24, 100 44 Stockholm, Sweden, email: loizosh@kth.se.

In short, pyspect compiles temporal logic specifications to reachability programs, checking consistency and correctness in the process. We have published the toolbox on GitHub[1] to invite and encourage the community to compare their reachability implementations through TLTs.

## II. PROBLEM FORMULATION

This section will first recall reachability analysis and temporal logic trees from [10] and [12]. Then, we use the preliminary material to state the main challenge we address with this work.

### A. System Dynamics

While pyspect is method-agnostic and is customizable with third-party solvers, the dynamical model used in reachability analysis is implementation-specific. However, for the sake of pedagogy, we start by considering systems with dynamics

$$\dot{z} = f(z, u) \quad (1)$$

where $z \in \mathbb{S} \subseteq \mathbb{R}^{n_z}$, $u \in \mathbb{U} \subseteq \mathbb{R}^{n_u}$ and $f$ is uniformly continuous, bounded, and Lipschitz continuous in $z$. We let the function space $U$ contain all control functions $T \mapsto \mathbb{U}$ over the time window $T = [t_0, t_f] \subseteq \mathbb{R}$. Starting from the initial state $z_0$, with $u(\cdot) \in U$, let $\zeta(t; z_0, t_0, u(\cdot)) \in \mathbb{S}$ be the state on the system's trajectory at $t \in T$, such that $\zeta(t_0) = z_0$. For simplicity, we may refer to the entire trajectory using $\zeta(\cdot)$.

Throughout the paper, we will mainly work with sets of states, $\mathbb{A} \subseteq \mathbb{S}$. We note that these sets can be represented in different ways for computational purposes, for example as level sets or hybrid zonotopes. We indicate representation through superscript, i.e. $\mathbb{A}^{(\text{HJ})}$ and $\mathbb{A}^{(\text{HZ})}$ are represented as a HJ level set and a hybrid zonotope, respectively.

### B. Temporal Logic

When defining objectives and requirements for a dynamical system, we can express them as specifications in temporal logic. While this paper primarily employs linear temporal logic (LTL), we will discuss how one can change the formal language in practice. To later demonstrate this point, we consider both standard LTL and a subset of LTL. Much of this subsection recalls from [2], [12].

Following [12], we first consider finite-trace LTL over a finite set of atomic propositions AP with syntax

$$\varphi ::= \top \mid p \mid \neg \varphi \mid \varphi_1 \wedge \varphi_2 \mid \varphi_1 \cup \varphi_2 \mid \bigcirc \varphi,$$

where $\top$ is "truth", $p \in \text{AP}$, $\neg$ and $\wedge$ denote negation and conjunction, as well as U and $\bigcirc$ are the "until" and "next" operators. With $\top, \neg$ and $\wedge$, we can derive all of propositional logic. With only U, we can further derive additional temporal operators such as $\Diamond \varphi = \top \cup \varphi$ (eventually) and $\square \varphi = \neg \Diamond \neg \varphi$ (always) [2]. With a labeling function relating states to propositions, $l : \mathbb{S} \to 2^{\text{AP}}$, we conversely define $\mathsf{M}(p) = \{z \in \mathbb{S} \mid p \in l(z)\}$ to map propositions to state sets. When working in continuous time, we will use a fragment of LTL without the $\bigcirc$ operator. We will consequently refer to this fragment as continuous-time LTL or $\text{LTL}^{\setminus \bigcirc}$.

We define a logic fragment $L$ as a syntactically restricted subset of a formal specification language (e.g., LTL) for which operational semantics can be assigned. To denote whether $\varphi$ is a well-formed formula, i.e. follows the syntax of $L$, we write $\varphi \in L$. For example, $\bigcirc \varphi \notin \text{LTL}^{\setminus \bigcirc}$. With this, we can move on to define temporal logic trees.

### C. Temporal Logic Trees

After writing a temporal logic formula $\varphi$, we construct a Temporal Logic Tree (TLT) to serve as a computational model for assessing its feasibility on a dynamical system [10], [12]. Structurally, the TLT recursively maps the syntactic structure of $\varphi$ into a tree of reachability- or set-based computations.

*Definition 2.1* (**Temporal Logic Tree**): A temporal logic tree is a tree for which each node is either a set node or an operator node; the root node and the leaf nodes are set nodes; if a set node is not a leaf node, its unique child is an operator node; the children of any operator node are set nodes.

Throughout the rest of this subsection, consider TLT $\mathfrak{T}_\varphi$ constructed from $\varphi$ as described by [12, Section IV]. Given a logic fragment $L$, a set of primitive TLTs $\mathsf{Q}_L$ defines an operational semantics for $L$ if it includes at minimum $\{\mathfrak{T}_\psi \mid \psi \text{ is atomic in } L\}$. This enables us to deconstruct a more complex TLT $\mathfrak{T}_\varphi$ into these primitives without explicitly evaluating the root node set, which enables static analysis

---

of the tree. To retrieve the root node set, we will need to compute reachable sets.

*Definition 2.2* (**Backward Reachable Set**): Given system (1), a time window $T = [t_0, t_f]$, a target set $\mathbb{T} \subseteq \mathbb{S}$, and a constraint set $\mathbb{C} \subseteq \mathbb{S}$, we define the backward reachable set as

$$\mathcal{R}_B(\mathbb{T}; \mathbb{C}) = \big\{ z \in \mathbb{S} \mid \exists u(\cdot) \in U,$$
$$\exists z_f \in \mathbb{T}, \zeta(t_f; z, t_0, u(\cdot)) = z_f,$$
$$\forall \tau \in T, \zeta(\tau; z, t_0, u(\cdot)) \in \mathbb{C} \big\}$$

where $\mathcal{R}_B(\mathbb{T}; \mathbb{C})$ is the set of states from which the system is able to reach the target set $\mathbb{T}$, while respecting the constraint set $\mathbb{C}$, within time window $T$.

In short, a primitive $\mathfrak{T}_\varphi$ with root node set $\Phi$, composed of branches $\mathfrak{T}_{\varphi_1}$ and $\mathfrak{T}_{\varphi_2}$ (with corresponding root node sets $\Phi_1$ and $\Phi_2$), defines its operator semantics as: $\Phi = \mathcal{R}_B(\Phi_2; \Phi_1)$ if $\varphi = \varphi_1 \cup \varphi_2$; $\Phi = \Phi_1 \cap \Phi_2$ if $\varphi = \varphi_1 \wedge \varphi_2$; and so on, depending on the connective. The robust control invariant set is normally used to operationalize $\varphi = \Box \varphi_1$ [12], however, we omit an exact definition to enable further discussion of $\Box$ later in the paper. Propositions are always formulated as set-membership conditions, meaning that $\Phi = \mathsf{M}(p)$ for a leaf node $\mathfrak{T}_p$, effectively $p \leftrightarrow z \in \Phi$. We *realize* $\mathfrak{T}_\varphi$ when we execute the underlying set and reachability operations to produce $\Phi$. Computationally, $\Phi$ is typically over- or under-approximated with respect to the true reachable set for $\varphi$. Hence, it becomes important to avoid approximation mismatch when TLTs are composed as this can falsely eliminate or include trajectories in the solution. Finally, we explicitly note that $\varphi$ is satisfiable if $\Phi$ is non-empty [10, Theorem V.1]. For further details, we direct interested readers to [12].

### D. Problem Statement

Current reachability frameworks often tightly couple the logic specification, solver implementation, and problem formulations, making it difficult to reuse specifications and compare methods. In this paper, we introduce pyspect, a toolbox that addresses this challenge.

First consider a verification problem encoded in the formula $\varphi \in L$. Using TLTs, we can automatically convert $\varphi$ to a reachability analysis program that solves the verification problem. In doing so, operators in $L$ correspond to certain computational procedures in the implementation $I$. For example, "$\vee$" will typically map to a procedure "union" in the implementation. On the other hand, if $\varphi$ contains operators of $L$ that are not supported by $I$ (e.g. $\varphi$ contains $\vee$ but $I$ does not have union), then the implementation cannot evaluate $\varphi$, leading to failure unless $\varphi$ is rewritten or $I$ is extended. We say that $\varphi$ and $I$ are *consistent* if all logic operators in $\varphi$ are supported in $I$ via a set of primitive TLTs, $\mathsf{Q}_L$, and *correct* if their operational semantics through $\mathsf{Q}_L$ produce a sound (i.e. without approximation mismatch) satisfaction set for $\varphi$. Evidently, $\varphi$, $L$, and $I$ are highly interdependent, making it difficult to evaluate two different implementations under the same specification. Today, this requires significant manual labor if $\varphi$ is complex. Furthermore, this also makes it difficult for downstream use of reachability analysis since it requires expertise knowledge of each particular implementation.

This issue becomes even more apparent when logically equivalent formulas can lead to different operational semantics. For example, the specification $\Box \varphi$ is logically equivalent to both $\neg \Diamond \neg \varphi$ and $\varphi \wedge \bigcirc(\varphi \wedge \bigcirc(...))$, but different reachability implementations may interpret these differently. In Section V, we will see Hamilton-Jacobi reachability deriving $\Box \varphi$ from the $\Diamond$-based form and Hybrid Zonotope-based reachability using one-step backward reachable sets for the $\bigcirc$-based form. These differences are not just syntactic, they affect both performance and correctness, making it important to explicitly choose the operational semantics.

In pyspect, we treat $\varphi$ and $I$ as being independently authored by users who may not be aware of one another's assumptions or constraints. We refer to these users as *end-users* (for $\varphi$) and *implementation developers* (for $I$). The core challenge is then to ensure that $\varphi$ ($\in L$) and $I$ remain consistent and correct. This is precisely the challenge pyspect addresses.

### III. Software Design

In this section we describe pyspect's software design using the code example in Listing 1. First, we overview our general design philosophy. Then, we discuss specific design aspects and relate them to the example.

#### A. Overview

To ground the design in practical terms, this subsection illustrates the two user roles introduced in the problem statement.

End-users write specifications such as $\varphi_{\text{task}}$ (lines 1-6), typically describing safety and liveness properties over system trajectories. These formulas can be written in a general fashion, using any connectives in first-order temporal logic. Additionally, end-users can state propositions symbolically using strings (e.g., 'goal') and define their set-meaning later. Internally, symbolic propositions are made explicit using a map $\mathsf{M}: \mathsf{AP} \to 2^\mathbb{S}$ which the end-user can update when needed (line 10).

Implementation developers provide the computational backend used to execute the reachability programs of TLTs. This is implemented with custom objects (line 13) that provide procedures such as union, intersect, and reach. They do not need to be aware of the full structure of the TLT or convert their inputs/outputs to a unified set representation. Implementations need only to provide procedures corresponding to the TLT primitives they are targeting. These are then automatically called by pyspect (line 18). This separation allows implementation developers to focus on optimizing low-level computations while remaining compatible with any logic fragment that requires those operations.

#### B. Constructing and Realizing TLTs

We have already introduced TLTs and their semantics in Section II.C. Here, we focus on how pyspect constructs, validates, and realizes TLTs from user-defined specifications and backend implementations. These steps are driven by TLTs' compositional structure and will depend on a set of primitive TLTs, $\mathsf{Q}_L = \{\mathfrak{T}_\cup, \mathfrak{T}_\wedge, \mathfrak{T}_\neg, ...\}$.

Listing 1: Code example that illustrates basic concepts in pyspect. The task is to verify an overtaking maneuver around a slow-moving vehicle characterized by $D$. $\varphi_{\text{env}}$ captures the scenario constraints, such as not colliding (not entering $D$). Then, we seek to reach the goal set $G$, i.e. being in front of the other vehicle and in the right lane ($y < 0$).

```
1  # WRITE SPECIFICATION
2  D    = {z ∈ S | 0 ≤ x − 16t ≤ 40, y < 0}
3  φ_env = AND(NOT( D ), ...)
```
▸ Propositions are often immediately associated to specific sets. Short-hand syntax allow $\varphi_{\text{env}}$ to infer proposition $d \leftrightarrow z \in D$.
▸ Specifications are stored as syntax trees:
   $\varphi_{\text{env}} = $ ('AND', ('NOT', d ), ...)
```
4
5  φ_task = UNTIL(φ_env, 'goal')
```
▸ Symbolic proposition: 'goal' $\in$ AP.
```
6
7  # CONSTRUCT TLT
8  TLT.select( Q_{LTL\O} )
9  G    = {z ∈ S | 40 < x − 16t, y < 0}
10 tree = TLT( φ_task , where={'goal': G })
```
▸ Select the set of primitives $Q_{\text{LTL}\backslash\bigcirc} = \{\mathfrak{T}_U, ...\}$ to construct TLTs, meaning tree is constructed only if $\varphi_{\text{task}} \in \text{LTL}^{\backslash\bigcirc}$.
▸ Update symbolic proposition: 'goal' $\leftrightarrow z \in G$.
```
11
12 # INITIALIZE IMPLEMENTATION
13 impl = HJImpl(dynamics...,
14                horizon...,
15                grid...)
```
▸ pyspect does not enforce what/how implementation-specific properties can be configured.
```
16
17 # RUN REACHABILITY
18 out = tree.realize(impl)
```
▸ Realizing tree involves verifying its consistency and calling the relevant procedures in impl to compute out $\subseteq \mathbb{S}^{(\text{HJ})}$.

Recall from the example that $\varphi_{\text{task}}$ is first written without reference to any specific logic fragment. At this stage, it is simply a syntactically valid formula in first-order temporal logic. Only once $Q_{\text{LTL}\backslash\bigcirc}$ is selected (line 8) can $\varphi_{\text{task}}$ be interpreted within $\text{LTL}^{\backslash\bigcirc}$. The call TLT($\varphi_{\text{task}}$) then attempts to construct $\mathfrak{T}_{\varphi_{\text{task}}}$ by recursively traversing its subformulas $\psi$. If $\psi = $ 'goal', we construct $\mathfrak{T}_\psi$ using $\mathsf{M}(\texttt{goal}) = G$, making $\mathfrak{T}_\psi$ a leaf node. This requires that 'goal' is included among the atomic propositions, which is ensured by "where={'goal': $G$}." More generally, if $\psi$ is atomic, then $\mathfrak{T}_\psi \in Q_{\text{LTL}\backslash\bigcirc}$. Otherwise, if $\psi \notin \text{LTL}^{\backslash\bigcirc}$, the construction fails as $\varphi_{\text{task}} \notin \text{LTL}^{\backslash\bigcirc}$. Thus, $Q_L$ defines the operational semantics of a logic fragment $L$, and thereby determines the set of formulas that can be constructed.

After construction, the TLT serves as an intermediate representation that guides evaluation: each operator node corresponds to a reachability or set operation delegated to the backend implementation. Realizing a TLT is therefore akin to executing a program over sets, where each instruction is defined by the logic's operational semantics. To do this, the end-user invokes realize(impl) (line 18), injecting an implementation that realizes the root node set of the tree by calling the corresponding set and reachability procedures in impl. At this stage, pyspect checks for semantic compatibility, i.e. that impl supports all operations required by $Q_{\text{LTL}\backslash\bigcirc}$. If all interface requirements are satisfied, the realization step computes the solution set out at the root node, representing the states from which the specification holds. In this workflow, primitive TLTs act as the interface between formal logic and reachability procedures, cleanly separating logic structure ($\text{LTL}^{\backslash\bigcirc}$), specification intent ($\varphi_{\text{task}}$), and numerical implementation (impl).

*C. Set Representation*

Another of our core design decisions addresses a concrete problem that the reachability community is currently facing. Namely, reachability frameworks are tightly coupled to their set representation, making it difficult to unify operations across different approaches. Particularly, across paradigms such as implicit and explicit representations. To overcome this, pyspect adopts the common design pattern of *dependency injection* to ensure a clean separation of concerns between implementation developers and end-users

Consider the definition of sets (lines 2 and 9). Here, even without the implementation instantiated, the end-user can describe and build them using "set-builder" primitives, such as BoundedSet(y=($y_{\text{lb}}, y_{\text{ub}}$)). These are lazily evaluated, meaning that the computation happens at a later stage. For example, $G$ is first represented implicitly as a function B that, given implementation $I$ and a proposition map M (internally managed to support referencing), computes $\mathsf{B}(I; \mathsf{M}) \subseteq \mathbb{S}^{(I)}$. The implementation serves as the so-called "injector", supplying the procedures needed to evaluate the set. Note that the resulting set is then in the same representation as used by $I$, which is exactly how TLTs are realized (line 18). As such, pyspect offers set-builder primitives for all standard set operations.

## IV. IMPLEMENTING REACHABILITY IN PYSPECT

This section lays out the necessary infrastructure to construct pyspect implementations. Specifically, we detail the implementation of two reachability frameworks, Hamilton-Jacobi (HJ) and Hybrid Zonotope (HZ).

*A. Hamilton-Jacobi Reachability*

The HJ framework poses reachability analysis as an optimal control problem which one can solve numerically with dynamic programming [14]. More specifically, we solve two variants of the Hamilton-Jacobi-Bellman (HJB) equation,

$$\underbrace{\partial_{t'} V(z, t') + \min_{u \in \mathbb{U}} \partial_z V(z, t') \cdot f(z, u)}_{\partial_\exists V(z,t')} = 0 \quad (2.1)$$

$$\underbrace{\partial_{t'} V(z, t') + \max_{u \in \mathbb{U}} \partial_z V(z, t') \cdot f(z, u)}_{\partial_\forall V(z,t')} = 0 \quad (2.2)$$

in a backwards fashion with $t' = t_f - t$ over $T = [t_0, t_f]$. Its viscosity solution, $V$, can be interpreted as an implicit surface function of which the sub-zero level set corresponds to the reachable set, i.e. $\{z \in \mathbb{S} \mid V(z, t') \leq 0\} \subseteq \mathbb{S}^{(\text{HJ})}$.

For brevity, we denote the left-hand sides of (2.1) and (2.2) as differential operators $\partial_\exists$ and $\partial_\forall$, corresponding to existential and universal quantification over control inputs. These reflect whether the reachability problem seeks $\exists u(\cdot)$ or $\forall u(\cdot)$ in Definition 2.2.

Since HJ reachability is formulated in continuous time, it only supports $\text{LTL}^{\setminus \bigcirc}$. To operationalize formulas defined with the U operator, we need the backward reachable set $\mathcal{R}_B(\mathbb{T}; \mathbb{C})$. For this, we are interested in the viscosity solution $V_\mathcal{R}$ of the constrained HJB equation

$$\max\{\partial_\exists V_\mathcal{R}(z, t'), -V_\mathbb{C}(z) - V_\mathcal{R}(z, t')\} = 0 \quad (3)$$

with initial condition $V_\mathcal{R}(z, 0) = V_\mathbb{T}(z)$. $V_\mathbb{T}$ and $V_\mathbb{C}$ must be Lipschitz continuous and may be constructed using the signed distance function to the set boundaries. Then, the sub-zero level set of $V_\mathcal{R}$ is the reachable set $\mathcal{R}_B(\mathbb{T}; \mathbb{C})$.

While this is sufficient for most temporal operators in $\text{LTL}^{\setminus \bigcirc}$, $\square$ requires a special procedure. To compute the root node set $\Phi$ of $\mathfrak{T}_{\square c}$, where $c \leftrightarrow z \in \mathbb{C}$, we first need the so-called avoid set $\mathcal{A}(\mathbb{T})$. Similar to $V_\mathcal{R}$, we solve for $V_\mathcal{A}$ in

$$\partial_\forall V_\mathcal{A}(z, t') = 0 \quad (4)$$

with initial condition $V_\mathcal{A}(z, 0) = V_\mathbb{T}(z)$. Using the avoid set, we get $\Phi = \mathcal{A}(\mathbb{C}^C)^C$. Tellingly, this form relates to the logical equivalence $\square \varphi = \neg \Diamond \neg \varphi$, except it is quantifying over universal control rather than existential which $\Diamond$ normally do. For more details on (3) and (4), we refer readers to [12], [14].

This approach is also called the level-set method, referring to the set representation. The major drawback with this representation is that all value functions are discretized over a grid, thus suffering from the curse of dimensionality. Typically, one can use HJ reachability for systems of up to 6 dimensions before it becomes intractable. There exists some techniques to address this, including system decomposition [15], warm-starting [16] and deep learning [17], enabling solving for more complex reachability problems. Still, this issue typically prohibits the adoption of HJ reachability.

Despite its drawbacks, the level-set method also offers several advantages. For one, level sets have no restrictions in the type of sets they can represent. For instance, the value functions can describe non-convex, and even fully disjoint, sets. Another benefit stems from their explicit, grid-based representation. Normal set operations are reduced to simple procedures that are easily parallelized. Consider two sets $\mathbb{A}$ and $\mathbb{B}$, and their corresponding value functions $V_\mathbb{A}$ and $V_\mathbb{B}$, respectively. Then, we have

$$V_{\mathbb{A}^C}(z) = -V_\mathbb{A}(z),$$
$$V_{\mathbb{A} \cap \mathbb{B}}(z) = \max\{V_\mathbb{A}(z), V_\mathbb{B}(z)\},$$
$$V_{\mathbb{A} \cup \mathbb{B}}(z) = \min\{V_\mathbb{A}(z), V_\mathbb{B}(z)\}.$$

These operations can introduce much more complexity in implicit representations.

### B. Hybrid Zonotope-based Reachability

The hybrid zonotope is a non-convex set representation constructed as the union of constrained zonotopes, which themselves generalize zonotopes by allowing linear equality constraints [18]. The representation is implicit, as generators and constraints encode the set without enumerating its elements, enabling compact storage and efficient operations.

*Definition 4.1* (**Hybrid Zonotope**): A set $\mathbb{A} \subset \mathbb{R}^{n_z}$ is a hybrid zonotope if there exist $G^c \in \mathbb{R}^{n_z \times n_g}, G^b \in \mathbb{R}^{n_z \times n_b}, A^c \in \mathbb{R}^{n_c \times n_g}, A^b \in \mathbb{R}^{n_c \times n_b}$, and $c \in \mathbb{R}^{n_z}, b \in \mathbb{R}^{n_c}$ such that

$$\mathbb{A} = \left\{ \begin{bmatrix} G^c & G^b \end{bmatrix} \begin{bmatrix} \xi^c \\ \xi^b \end{bmatrix} + c \;\middle|\; \begin{matrix} \begin{bmatrix} \xi^c \\ \xi^b \end{bmatrix} \in \mathcal{B}_\infty^{n_g} \times \{-1, 1\}^{n_b}, \\ \begin{bmatrix} A^c & A^b \end{bmatrix} \begin{bmatrix} \xi^c \\ \xi^b \end{bmatrix} = b \end{matrix} \right\},$$

with short-hand notation $\mathbb{A} = \langle G_c, G_b, c, A_c, A_b, b \rangle$.

Hybrid zonotopes enable an effective approach for reachability analysis in discrete-time hybrid systems. They are closed under many common set operations, including linear mapping, Minkowski sum, intersection with halfspaces and more, making them compositional and amenable to forward propagation. To support reachability analysis with hybrid zonotopes, we work with a discrete-time reformulation of (1). Specifically, we consider the linear, discrete-time system

$$z_{t+1} = A z_t + B u_t, \quad (5)$$

where $z_t \in \mathbb{R}^{n_z}$ is the system state, $u_t \in \mathbb{U}$ is the control input, and $A \in \mathbb{R}^{n_z \times n_z}, B \in \mathbb{R}^{n_z \times n_u}$ are the system and input matrices. We assume the system evolves deterministically given $z_t$ and $u_t$, and define a control policy $\mu = u_0, u_1, \ldots$ from the space of admissible policies $\mathcal{M}$. To follow the system evolution, we introduce the predecessor set (also known as 1-step backward reachable set) similar to $\mathcal{R}_B$.

*Definition 4.2* (**Predecessor Set**): Given system (5), a target set $\mathbb{T} \subseteq \mathbb{S}$, and a constraint set $\mathbb{C} \subseteq \mathbb{S}$, we define the predecessor set as

$$\mathcal{P}(\mathbb{T}; \mathbb{C}) = \{z \in \mathbb{C} \mid \exists u \in \mathbb{U}, Az + Bu \in \mathbb{T}\},$$

where $\mathcal{P}(\mathbb{T}; \mathbb{C})$ is the set of states in $\mathbb{C}$ that can reach $\mathbb{T}$ in one discrete time step.

When analyzing system (5) using hybrid zonotopes, we perform reachability analysis over a finite time horizon by recursively applying predecessor operations. Namely, over $N$ steps, the backward reachable set is $\mathcal{R}_B(\mathbb{T}; \mathbb{C}) = \mathcal{P}(\mathcal{P}(\ldots; \mathbb{C}); \mathbb{C})$ with the $N^{\text{th}}$ predecessor calling $\mathcal{P}(\mathbb{T}; \mathbb{C})$. For standard LTL, the predecessor set operationalizes the $\bigcirc$ operator. This allows for $\square \varphi = \varphi \wedge \bigcirc(\varphi \wedge \bigcirc(\ldots))$ which, in contrast to HJ, computes $\Phi = \cap_{i=0}^N \mathcal{R}_B(\mathbb{S}; \mathbb{C})$ as the root node set for $\mathfrak{T}_{\square c}$. The U operator is similarly operationalized through discrete steps with $\bigcirc$. Specifically, we implement the hybrid zonotope-based reachability analysis as described in [13, Section IV.B].

In addition to the predecessor set, hybrid zonotopes support efficient implementations of the standard set operations relevant for TLTs (complement, intersection, union). Rather than relying on grid-based value functions or point-wise set evaluations, these operations are carried out directly on the generator and constraint structure of the hybrid zonotope. For example, through a number of matrix concatenations to compute Minkowski sum and intersections [18, Section 3.2].

Although these methods are efficient, hybrid zonotopes may grow in both binary and continuous dimensions over time, increasing complexity and potentially adding cost to downstream applications. To address this, the original authors propose techniques for pruning redundant generators and using binary tree decompositions [18]. These techniques preserve correctness while being closed in the representation. However, practical use still requires managing rapid growth in generator dimensions, often requiring partial enumeration and solving mixed integer linear programs. Nevertheless, hybrid zonotopes preserve a symbolic and closed structure under set operations, enabling efficient and scalable reachability analysis over discrete time horizons.

*C. Realizing TLTs in HJ and HZ*

While the previous subsections detailed how HJ and HZ reachability compute specific types of sets, these methods become usable in pyspect only once they are linked to the supported temporal logic $L$, which is done through TLT primitives $Q_L$. The primitives do not require implementations to be deeply integrated in pyspect, rather only define the interface for how to call procedures in the implementation, such as reach, intersect, union, or complement. This enables pyspect to construct a program of nested procedure calls that computes the output set. For a given TLT $\mathfrak{T}_\varphi$, this program is the set-builder function B introduced in Section III.C, representing the TLT's root node set. Thus, any compatible implementation $I$ can realize $\mathfrak{T}_\varphi$ by evaluating $B(I)$, which triggers the various max/min operations in HJ or matrix concatenations in HZ, to produce the root node set $\Phi^{(I)}$.

## V. EXAMPLES

In this section, we illustrate how pyspect can define and evaluate a temporal logic specification across different reachability frameworks. Specifically, we will contrast two formulas, $\varphi_1 = \square \psi$ and $\varphi_2 = \square \Diamond \psi$, highlighting the issue mentioned in Section II.D and the benefit of using pyspect. Finally, we end the section with a more complex case study.

For the two main examples, we consider simple double integrator dynamics modelling a vehicle's longitudinal motion: $\dot{x} = v$, $\dot{v} = u$. The system state $[x, v]^T \in [-100, 100] \times [-10, 10]$ encodes the vehicle's position $x$ and velocity $v$ along a straight lane. The system is controlled via acceleration $u \in \mathbb{U}$, subject to a control bound $|u| \leq a_{\max} = 1.0 \frac{m}{s^2}$. Next, we consider both continuous-time and discrete-time variants of the system, following (1) and (5), over a 40 s time horizon. For (5), we define $A$ and $B$ derived via zero-order hold over a sampling time $\Delta t = 0.5$ s. Finally, we consider the proposition $s$ for when the vehicle is within the strip $-50 \leq x \leq 50$. For both formulas, we set $\psi = p$ such that $M(p) = \{[x, v]^T \mid x \in [-50, 50]\}$ is the root node set of the subtree $\mathfrak{T}_\psi$. With this, we can construct $\mathfrak{T}_\varphi$ and realize it using compatible backend implementations, in this case with HJ and HZ reachability analysis.

All computations in this section were performed on a workstation equipped with an Intel Core i9-12900K CPU and an NVIDIA RTX 3090 GPU (24 GB), running Ubuntu 22.04 with CUDA 12 and NVIDIA driver 560.

*A. Specification:* $\square \psi$

The $\square$ operator can be interpreted in two ways. With the HJ framework, $\square \psi$ is evaluated via $\neg \Diamond \neg \psi$ solving a single backward reachable (avoid) set. As a result, the corresponding TLT (see Fig. 1) contains only a small number of nodes, and the approximation behavior is conservative, yielding an under-approximation of the safe set. In contrast, the HZ implementation uses the alternative fixed-point formulation $\psi \wedge \bigcirc(\psi \wedge \bigcirc(...))$, operationalized as an iterated sequence of predecessor operations. This approach aligns with the strengths of zonotopic reachability, which supports symbolic, discrete-time propagation with efficient set operations. However, it leads to a deeper tree and the discrete-time nature of the system also introduces differences in how the reachable sets evolve over time, e.g. control is fixed during each time step due to zero-order hold.

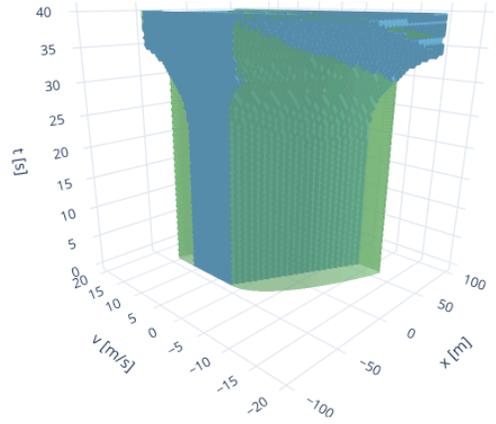

Fig. 2: Satisfaction sets $\Phi_1^{(HJ)}$ (green) and $\Phi_1^{(HZ)}$ (blue) for $\varphi_1 = \square \psi$.

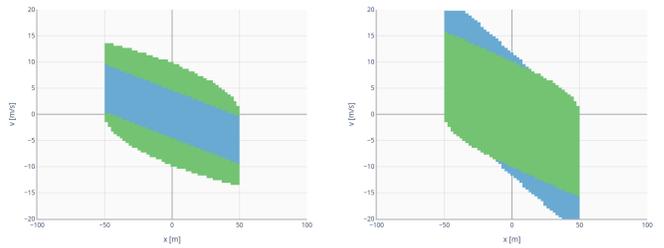

Fig. 3: Two snapshots of satisfaction sets $\Phi_1^{(HJ)}$ (green) and $\Phi_1^{(HZ)}$ (blue), at times $t = 0$ s and $t = 35$ s, respectively.

Despite these differences, both realizations are derived from the same temporal logic formula $\varphi_1$, constructing the same TLT in pyspect. The specification $\varphi_1 = \square \psi$ is written once by the end-user, then validated, and realized using the backend of choice. Each backend only implements the set operations to the relevant TLT primitives in $Q_{LTL \setminus \bigcirc}$ and $Q_{LTL}$, respectively, without access to $\varphi_1$ or knowing which logic fragment is used. This separation of concerns allows users to switch between backends or compare them side-by-side without refactoring their original logic. We illustrate this with Fig. 2 and 3, showing the root node sets $\Phi_1^{(HJ)}$ and $\Phi_1^{(HZ)}$ of $\mathfrak{T}_{\varphi_1}$, when using the corresponding implementations for HJ and HZ. As outlined in Section II.C, satisfiability corresponds to non-emptiness of the root node set, i.e. $\Phi_1 \neq \emptyset$. In Fig. 2, we see the root node sets as they change over time. As we are further from the time horizon $t = 40$, the satisfaction sets become smaller as the vehicle must conform to $\varphi_1$ for longer. In Fig. 3, we see two snapshots of the same satisfaction sets, at $t = 0$ and $t = 35$, respectively. We see that they produce similar solutions, noting that these methods under-approximate the true reachable set to various degrees at different times. The HJ approach required approximately 0.24 s to solve on a $91 \times 91$ grid, while the HZ approach completed significantly faster in under 0.03 s, yielding a hybrid zonotope with 559 continuous generators and 478 equality constraints. These results highlight how the same formal specification $\varphi_1$ can be realized with different performance and structural tradeoffs, all within `pyspect`'s modular interface.

### B. Specification: $\Box \Diamond \psi$

A key feature of `pyspect` is its automatic validation of approximation direction during TLT construction. When composing logic formulas for reachability analysis, it is often easy to overlook mismatched approximation types, such as combining an over-approximated reach set with an under-approximated invariant. These mismatches can silently violate the intended semantics of the specification and lead to unsound results. `pyspect` statically analyzes the TLT to ensure consistency between approximation direction and logical interpretation. The specification $\varphi_2 = \Box \Diamond \psi$, meaning "it is always possible to eventually satisfy $\psi$", highlights this feature for HJ and HZ.

In our case, $\Box$ and $\Diamond$ are normally under-approximated. With the HJ implementation, both negations and the $\Diamond$ embedded in $\Box$ are well-supported [12]. Consequently, $\mathfrak{T}_{\varphi_2}$ yields the satisfaction set in Fig. 4, showing how it changes over time. Notably, this formula is more permissive as it covers the full state space during $t \in [0, 23]$. Realizing $\mathfrak{T}_{\varphi_2}$ with HJ, i.e. computing $\Phi_2^{(\mathrm{HJ})}$, takes 1.44 s on the same $91 \times 91$ grid. In contrast, HZ interprets $\varphi_2$ as a fixed-point over intersections of predecessor sets at each step, $\Diamond \psi \wedge \bigcirc(\Diamond \psi \wedge \bigcirc(...))$. To reduce generator complexity, it is normal for the predecessor set to under-approximate its solution. Intersecting under-approximated sets can eliminate (or fail to eliminate) trajectories that conform to the true semantics of the logic, thus producing unsafe under-approximations [12]. Hence, at construction, `pyspect` detects, reports, and (by default) rejects the realization of $\mathfrak{T}_{\varphi_2}$ for HZ. This does not reflect a limitation of the HZ framework, but rather highlights how different implementations lead to different operational semantics when realizing the same logical specification. The same formula may be supported in one backend but require reformulation or alternative semantics in another. By validating such cases at construction, `pyspect` helps end-users avoid subtle semantic mismatches and ensures that each realization remains sound with respect to its specification.

### C. Case Study: Overtaking Scenario

To showcase `pyspect` on a more complex task, we consider again Listing 1. The goal is to complete an overtaking maneuver, returning to the right lane ahead of a slow-moving vehicle. Following the workflow in Section III, the specifi-cation $\varphi_{\mathrm{task}} = \varphi_{\mathrm{env}} \, \mathsf{U} \, \mathtt{'goal'}$, which encodes lane-keeping and safety requirements in $\varphi_{\mathrm{env}}$, is expressed independently of the backend implementation and compiled into a temporal logic tree. Since the vehicle must remain within the drivable corridor and avoid collision with the observed vehicle in front, we have $\varphi_{\mathrm{env}} = \neg d \wedge (p_{\mathrm{right}} \vee p_{\mathrm{left}} \vee p_{\mathrm{switch}})$ where $d \leftrightarrow z \in D$, and $p_{\mathrm{right}}, p_{\mathrm{left}}, p_{\mathrm{switch}}$ are propositions for being in the right lane, left lane, and switching lane, respectively.

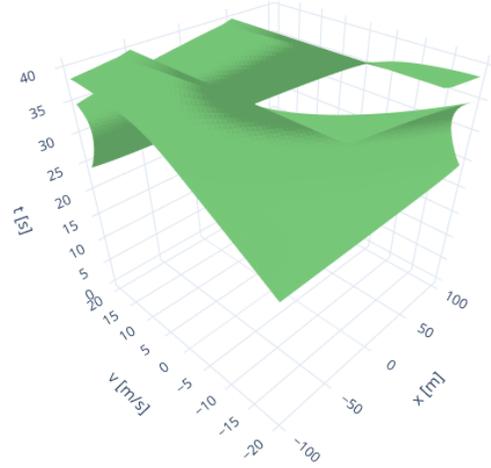

Fig. 4: Satisfaction set $\Phi_2^{(\mathrm{HJ})}$ over time for $\varphi_2 = \Box \Diamond \psi$. During $t \in [0, 23]$, the full state space is admissible. However, within the last 15 s, some states outside the strip and near the edges within the strip does not satisfy $\varphi_2$.

Unlike previous examples, which focused on 2D double-integrator dynamics, here we use the 4D kinematic bicycle model. Specifically, the state $z = [x, y, \theta, v]^T$ evolves with

$$f(z, u) = \begin{bmatrix} v \cos \theta \\ v \sin \theta \\ \frac{v \tan \delta}{L} \\ a \end{bmatrix}$$

where $L$ is the wheelbase and the control input includes acceleration and steering angle, $u = [a, \delta]^T$. In contrast to the double integrator examples, this scenario requires reasoning about a higher-dimensional, nonlinear system where set boundaries are not necessarily aligned with the coordinate axes. Such specifications are typical in autonomous driving contexts, illustrating the need for flexible verification tools.

As shown in Listing 1, we build $\mathfrak{T}_{\varphi_{\mathrm{task}}}$ as an intermediate step before committing to a set representation or numerical scheme. Since HJ directly supports continuous-time nonlinear systems, we select it as the backend without altering the original logic specification. Thus, when realizing $\mathfrak{T}_{\varphi_{\mathrm{task}}}$, we apply level-set methods to compute the satisfaction set $\Phi_{\mathrm{task}}^{(\mathrm{HJ})}$, see Fig. 5. The computation took 7.18 s on a $201 \times 11 \times 13 \times 31$ grid. Fig. 6 shows annotated snapshots of the same set, illustrating that states in $\Phi_{\mathrm{task}}^{(\mathrm{HJ})}$ guarantee that the vehicle can complete the overtaking maneuver within the horizon. Taken together, this case study highlights a key feature of `pyspect`: from specification to realization, it enables formal verification of complex scenarios without requiring significant manual effort or deep domain expertise.

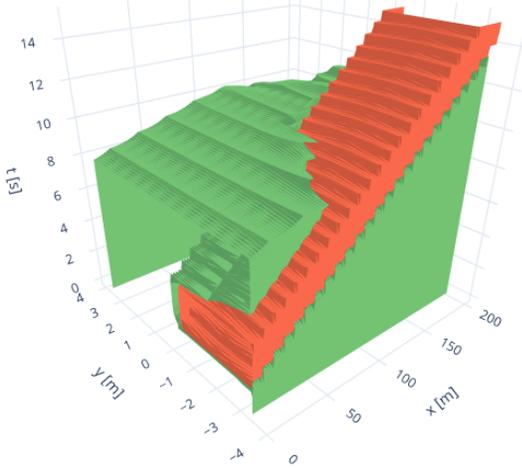

Fig. 5: Satisfaction set $\Phi_{\text{task}}^{(\text{HJ})}$ (green) projected over spatial coordinates $(x, y)$ and time $(t)$. This visualization highlights how the feasible overtaking trajectories align with lane constraints and avoiding $D$ (red).

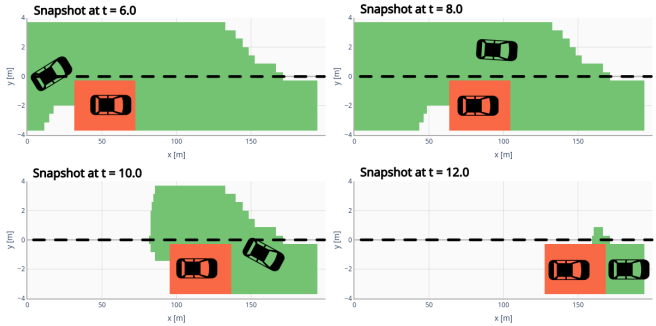

Fig. 6: Satisfaction set $\Phi_{\text{task}}^{(\text{HJ})}$ (green) for the overtaking case study, projected over spatial and temporal dimensions. $D$ (red) indicates the slow-moving vehicle. The satisfaction set represents states from which the overtaking maneuver can be completed within the time horizon.

## VI. Conclusion

In this paper, we presented `pyspect`, a method-agnostic Python toolbox for defining and solving reachability problems using Temporal Logic Trees (TLTs). By decoupling the specification language, implementation backend, and problem definition, `pyspect` enables modular and reusable formulations of formal verification tasks. Through TLTs, we provide a clear interface between temporal logic specifications and set-based reachability computations, making it possible to analyze, compare, and deploy different solvers while preserving consistency and correctness guarantees. In future work, we plan to use `pyspect` to conduct more comprehensive comparisons across reachability frameworks. We invite the community to contribute by submitting new implementations and logic fragments, which we hope enables side-by-side benchmarking, interoperability, and formal specification reuse in reachability applications.